\documentstyle[multicol,epsfig,aps,prl]{revtex}

\begin{document}


\title{Phase transitions in two dimensions - 
  the case of Sn adsorbed on Ge(111) surfaces}
\newcommand{\thispaper}{O. Bunk {\it et al.\@},\\
  Phase transitions in two dimensions -\\
  the case of Sn adsorbed on Ge(111) surfaces}

\author{O. Bunk,\cite{author_email} 
  J. H. Zeysing, G. Falkenberg, and R. L. Johnson} 
\address{II. Institut f\"ur
  Experimentalphysik, Universit\"at Hamburg, Luruper Chaussee 149, 
  D-22761 Hamburg, Germany}

\author{M. Nielsen, M. M. Nielsen, and R. Feidenhans'l}
\address{Condensed Matter Physics and Chemistry Department, 
  Ris\o\ National Laboratory, DK-4000 Roskilde, Denmark}

\date{March 9, 1999}

\maketitle

\begin{abstract}
Accurate atomic coordinates of the room-temperature 
($\sqrt{3}\times\!\sqrt{3}$)R30$^\circ$ and low-temperature 
(3$\times$3) phases of 1/3~ML Sn on Ge(111) have been established 
by grazing-incidence x-ray diffraction with synchrotron radiation. 
The Sn atoms are 
located solely at T$_4$-sites in the ($\sqrt{3}\times\!\sqrt{3}$)R30$^\circ$ 
structure. 
In the low temperature phase one of the three Sn atoms per 
(3$\times$3) unit cell is displaced outwards 
by 0.26$\pm 0.04$~\AA{} relative to the other two. 
This displacement is accompanied by an increase in the first 
to second double-layer spacing in the Ge substrate. 
\end{abstract}



\begin{multicols}{2}


Phase transitions at surfaces have aroused considerable interest among both 
theoreticians and experimentalists because they impact a wide variety of 
fields ranging from industrially important catalytic processes to providing 
insights into phenomena observed in cuprate superconductors. The suggestion 
that a commensurate charge density wave can form in 
Pb \cite{carpinelli_3x3pb_nature381} and 
Sn \cite{carpinelli_3x3sn_prl79,goldoni_3x3sn_prl79} overlayers on 
Ge(111) demonstrated the importance of such simple model systems as testing 
grounds for modern theories \cite{scandalo_surfsci402}. 
Upon cooling both of these adsorbate systems undergo a structural phase 
transition from a surface reconstruction with a 
($\sqrt{3}\times\!\sqrt{3}$)R30$^\circ$ periodicity at room temperature to a 
(3$\times$3) periodicity at low temperatures. 
For the Pb/Ge(111) system the (3$\times$3) structure is accompanied by a small
gap opening up in the electronic band structure, indicative of a 
metal-insulator transition.
The picture of a symmetry breaking transition was seriously questioned
in two recent papers, which reported almost identical electronic structures 
for both phases \cite{uhrberg_3x3sn_prl81,avila_prl82_3x3sn}. 
The transition was proposed to be of order/disorder
type with the Sn atoms fluctuating between two positions at 
room temperature, but freezing into an ordered (3$\times$3) structure 
at low temperature with an outwards displacement of every third Sn atom 
\cite{avila_prl82_3x3sn}. 
This throws into question the generally accepted T$_4$ model for the 
($\sqrt{3}\times\!\sqrt{3}$)R30$^\circ$ structure, in
which the adsorbate atom is located at a single threefold hollow site above a 
second layer Ge atom \cite{pedersen_sqrt3sn_surfsci189190} as shown in 
Fig.~\ref{fig:models}. To add to the confusion
the postulated Sn atom displacement in the (3$\times$3) structure 
was not found in a recent surface x-ray diffraction (SXRD) 
study \cite{baddorf_3x3sn_prb57}. 

Challenged by the discrepancies between the electronic and structural studies
performed so far, we decided to undertake a thorough 
investigation using surface x-ray diffraction to determine the geometrical 
structure of the Ge(111)-Sn system both at room and low temperature and 
by comparison to determine unambiguously the nature of the 
phase transition.


The samples were prepared in an ultra high vacuum (UHV) system equipped with 
reflection high energy electron diffraction,
low energy electron diffraction and a scanning tunneling microscope 
(STM). The substrates were cleaned using the standard procedure of repeated  
sputter-anneal cycles (500~eV Ar$^+$ ions, 450$^\circ$C) until good 
c(2$\times$8) diffraction patterns were observed. 
Tin was deposited from a calibrated effusion cell with the Ge(111) substrate 
held at room temperature; afterwards the sample was annealed to 
$\sim$~150$^\circ$C. This procedure yielded a 
Ge(111)-($\sqrt{3}\times\!\sqrt{3}$)R30$^\circ$-Sn 
reconstruction with a tin coverage very close to the ideal value of 
1/3~ML. 
STM measurements revealed well-ordered domains extending over 
$\sim$~400-600~\AA{} with a typical defect density of $\sim$4~\%, 
and the absence of the low coverage ``mosaic'' phase 
with a mixture of Sn and Ge adatoms. 
The sample was then transferred in a portable UHV chamber equipped with 
a closed-cycle sample cooling system to the BW2 wiggler 
beamline at HASYLAB for the x-ray diffraction measurements. 
The x-ray photon energy was set to 8.8~keV and a glancing angle of 
incidence to 0.8$^\circ$ was used (i.e.\ above the critical angle to reduce 
the uncertainties in the measured intensities arising from mechanical 
displacements). 
A data-set consisting of 35 symmetry inequivalent in-plane reflections, 
250 reflections along 14 fractional order rods and 62 reflections along 
three crystal truncation rods (CTRs) was recorded for the 
($\sqrt{3}\times\!\sqrt{3}$)R30$^\circ$ structure determination. 
After completing the room temperature measurements the sample was cooled until
the temperature of the sample holder reached 20~K. 
For the low temperature (3$\times$3) phase 278 reflections along 
17 fractional order rods and 19 reflections along one CTR were recorded. 
The three rods specific 
to the (3$\times$3) structure were rather weak and to optimize the 
signal to background ratio these were measured with the angle of incidence 
set to the critical angle and the data were scaled accordingly. 
The condition of the sample was checked by measuring a standard reflection 
at hourly intervals. 
The integrated 
intensities were corrected for the Lorentz factor, polarization factor, 
active sample area and the rod interception appropriate for the 
z-axis geometry \cite{vlieg_correction_factors}. 
The width of the fractional 
order reflections from the ($\sqrt{3}\times\!\sqrt{3}$)R30$^\circ$ phase 
corresponded to domains about ~500~\AA{} in diameter and this value did not 
change upon cooling. The reflections specific to the 
(3$\times$3) structure were considerably broader corresponding to 
an average domain size of only $\sim$~120~\AA{}. 
This indicates that cooling does not change the basic structure of the 
surface reconstruction, but it is modified by the superposition of a less 
well-correlated distortion. 
In the following we use the conventional surface coordinate system with 
${\bf a} = 1/2\,[10\overline{1}]_{\mbox{cubic}}$, 
${\bf b} = 1/2\,[\overline{1}10]_{\mbox{cubic}}$ and
${\bf c} = 1/3\,[111]_{\mbox{cubic}}$. The cubic coordinates are 
in units of the germanium lattice constant (5.66~\AA{} at 300~K).


A subset of the measured surface diffraction data is shown in 
Fig.~\ref{fig:rods}. 
The rods for the
($\sqrt{3}\times\!\sqrt{3}$)R30$^\circ$ and the (3$\times$3) phase are very
similar, but a careful inspection shows that there are important differences. 
Some of the rods are basically identical, as can be seen for the 
(2/3,\,5/3) or (4/3,\,1/3) rods, 
whereas for the (2/3,\,8/3) or (7/3,\,1/3) rods the intensities from the 
(3$\times$3) structure are 
significantly higher than those of the ($\sqrt{3}\times\!\sqrt{3}$)R30$^\circ$ 
structure. 
These differences are due to solely the changes in the atomic positions 
as a function of temperature.

In order to pinpoint the differences we first determined the
atomic positions of the 
($\sqrt{3}\times\!\sqrt{3}$)R30$^\circ$ structure using a least-squares 
refinement procedure. The atomic coordinates are given in 
Table~\ref{tab:abs_pos} and a ball and stick model of the structure 
with the displacements relative to bulk-like positions is shown in 
Fig.~\ref{fig:models}b. The Ge-Ge bond lengths deviate less than 3~\% from 
the bulk value of 2.45~\AA{}. The Ge-Sn bonds with 2.83$\pm 0.02$~\AA{} are 
slightly larger than the sum of the tetrahedral covalent radii for 
germanium and white 
tin (2.74~\AA{}/2.82~\AA{}) and larger than the value expected 
for grey tin ($\alpha$-Sn) and germanium (2.63~\AA{}). 
The Sn bond angle is 82.0$^\circ$. 
The in-plane displacement of the first layer Ge atoms of 0.17~\AA{} 
is significantly larger than the value
of 0.05~\AA{} given in Ref.~\cite{baddorf_3x3sn_prb57} and indicates that
the earlier analysis was based on a too limited dataset. 
The results of a Keating energy minimization are incompatible with 
the smaller value so we are forced to 
conclude that the analysis presented in Ref.~\cite{baddorf_3x3sn_prb57} 
is incorrect.
As shown in Fig.~\ref{fig:rods} the curves calculated using our structural 
model reproduce the experimental data extremely well
and this is confirmed by the reduced $\chi^2$ value 
of 1.6. 

Next, we determined the atomic coordinates of the low-temperature 
(3$\times$3) reconstruction and obtained the values
listed in Table~\ref{tab:abs_pos}. The differences between the
(3$\times$3) structure and the ($\sqrt{3}\times\!\sqrt{3}$)R30$^\circ$ 
structure are illustrated in  Fig.~\ref{fig:models}c. 
There are several important features to be noted: 
(i) One Sn atom is displaced out of the surface plane by 0.29~\AA{}; 
this Sn atom is at a vertical position 0.26~\AA{} higher than the average 
position of the two lower Sn atoms. 
(ii) The three nearest-neighbor Ge atoms partially follow this relaxation, 
mainly in the $z$-direction and not in-plane, contrary to what was 
reported in Ref.~\cite{baddorf_3x3sn_prb57}. 
(iii) The average layer spacing between the first and second Ge double-layer
is expanded relative to the room temperature phase. 
For the room temperature phase this distance  
is 1.004 and from the second to the third double-layer 0.993 in 
lattice coordinates, i.e.\ an expansion and a contraction relative to the 
bulk value of 1.000. However, for the low-temperature phase the first to 
second double-layer distance is 1.026 and the second to third layer distance
is 1.002, i.e.\ a considerable expansion in the upper two double-layers. 
(iv) The outwards displaced Sn atom has a very anisotropic atomic displacement 
parameter (adp) with an amplitude ten times larger in the $z$-direction than 
in-plane. 
This means that either the atom is performing a very anisotropic motion 
with a large amplitude or, as is more likely at low temperatures, there is 
some disorder in the $z$-position of this atom. 
The adp's for the nearest-neighbor Ge atoms 
are also larger than at room temperature again indicative of disorder. 
This is not 
surprising since the position of these atoms must at least partially follow 
the Sn atoms. 
The reduced $\chi^2$ for the 
low temperature data is 1.3; a subset of the fractional order rods 
is shown in Fig.~\ref{fig:rods}. 

A trial using a single isotropic adp for all Sn atoms resulted in an increased 
outward displacement of one Sn atom and an inward displacement of 
the two other Sn atoms with a total height 
difference of $\sim$~0.45~\AA{} between the Sn atoms. However, 
the three rods specific to the (3$\times$3) structure were 
not adequately described by this model. 

Several tests were performed to ensure that the features of the 
low-temperature phase were real and not caused by artifacts or 
local minima in the $\chi^2$ minimization. 
First, to check the sensitivity of 
the structure determination to changes in the relative weight of reflections 
we set the error bars on all measured reflections equal and re-optimized 
every parameter.
Although there were some minor differences the main features of the 
outward displacement and highly anisotropic adp for one Sn atom remained. 
In the second test we optimized the Ge positions in the third to sixth layers 
using a Keating model to
minimize the elastic strain energy \cite{keating}. All deviations were less 
than 0.06~\AA{}, so we can rule out the possibility that the good 
agreement between the measured and calculated intensities arises from 
unphysical atomic displacements in the substrate. In the third test 
we checked whether the low temperature displacements are dependent on 
the weak rods specific to the (3$\times$3) periodicity, which have 
larger relative uncertainties than the other rods. By excluding these rods 
from the data analysis and re-optimizing the parameters only minor changes, 
typically $<$ 0.03~\AA{}, occurred. 
From these checks we are convinced that our data analysis has revealed the 
intrinsic features of the low-temperature (3$\times$3) phase.

Now we can address the classification of the transition between the
($\sqrt{3}\times\!\sqrt{3}$)R30$^\circ$ and the (3$\times$3) phase  in 
more detail. Recently, it was proposed to be an order/disorder
transition \cite{uhrberg_3x3sn_prl81,avila_prl82_3x3sn}. 
This would require two different sites for the Sn atom 
with a height difference of about 0.26~\AA{} even at room temperature. 
However, if this were the case, there would be no difference between the 
($\sqrt{3}\times\!\sqrt{3}$)R30$^\circ$  specific rods in the 
($\sqrt{3}\times\!\sqrt{3}$)R30$^\circ$ and (3$\times$3) phase apart from 
the thermal motion effects affecting all rods, in contrast 
to what we observed experimentally as shown in Fig.~\ref{fig:rods}. 
To quantify this, we used the (3$\times$3) 
low temperature structure and optimized the displacements using the room 
temperature data. This gave a more isotropic 
adp for the outwards displaced Sn atom and a reduction of the outwards 
displacement to 0.07~\AA{}. 
The reduced $\chi^2$ in this test increased compared to the best fit for the
($\sqrt{3}\times\!\sqrt{3}$)R30$^\circ$ structure from 1.6 to 1.7  
due to the 
increase in the number of free parameters. Hence, we can conclude that 
if there is more than one site for the Sn atoms in the 
($\sqrt{3}\times\!\sqrt{3}$)R30$^\circ$ structure 
the height difference is much less than that observed in
the (3$\times$3) phase. The fact that the adp of the surface layer Ge atoms
are similar in both phases is strong evidence
against an order/disorder phase transition. At low temperatures one would 
normally expect both reduced thermal motion and disorder. 
The experimentally observed lattice distortion is reminiscent of a 
pseudo-Jahn-Teller-effect \cite{pick_surfscirep12} in which the energy of 
the system is lowered by a spontaneous symmetry-reducing displacement.


In summary, by performing a detailed analysis of comprehensive sets of 
x-ray diffraction data we have established definitive structural models 
for both the
room-temperature Ge(111)-($\sqrt{3}\times\!\sqrt{3}$)R30$^\circ$-Sn 
and low-temperature Ge(111)-(3$\times$3)-Sn surface reconstructions. 
The atomic coordinates are given in Table~\ref{tab:abs_pos}. 
The major feature of the (3$\times$3) structure is the
outward displacement of one Sn atom by 0.26$\pm 0.04$~\AA{} with respect 
to the average position of the other two Sn atoms per unit cell. 
The three nearest-neighbor Ge atoms bonding to the displaced Sn atom are 
also displaced outwards.
In addition there is an increase in the average layer spacing between the 
first to second Ge double-layers compared to the 
($\sqrt{3}\times\!\sqrt{3}$)R30$^\circ$ phase. 
We have shown that the phase transition from the 
($\sqrt{3}\times\!\sqrt{3}$)R30$^\circ$ 
to the (3$\times$3) phase is not an order/disorder transition.
We hope that the detailed structural information presented here 
will provide the foundation 
for a better theoretical understanding of this interesting model system. 

{\it Note added:} 
Zhang {\it et al.\@} have re-analyzed their previously published SXRD data 
for the Ge(111)/Sn phases \cite{baddorf_3x3sn_prb57} in conjunction with 
IV-LEED data \cite{zhang_ge111sn_prb:1999} and find 
an outward displacement of 0.37~\AA{} for one Sn atom. 
However, the displacements in the substrate differ from the values reported 
in this letter. 
A comparable outward displacement of a Pb atom by $\sim$0.4~\AA{} was 
described in a recent publication on the 
Ge(111)-(3$\times$3)-Pb structure by Mascaraque {\it et al.\@} 
\cite{mascaraque_ge111pb_prl:1999}. 


We thank the staff of HASYLAB for their technical assistance. 
Financial support from the Danish Research Council through Dansync, 
the Bundesministerium f\"ur Bildung, Wissenschaft, 
Forschung und Technologie (BMBF) under project no. 05~SB8~GUA6 and the 
Volks\-wagen Stiftung is gratefully acknowledged. 


\newpage


\begin{figure}[htb]
  \caption{(a) Top view of the Ge(111)-{\hskip 0pt}(3$\times$3)-Sn 
    reconstruction. The dashed and dotted lines mark the
    (3$\times$3) and ($\sqrt{3}\times\!\sqrt{3}$)R30$^\circ$ 
    unit-cells. The solid line marks the cut for the 
    side views shown in (b) and (c). 
    (b) Side view of the 
    Ge(111)-{\hskip 0pt}($\sqrt{3}\times\!\sqrt{3}$)R30$^\circ$-Sn 
    reconstruction. The displacements relative to bulk-like positions 
    in [1$\overline{\mbox{1}}$0] and [001] directions are given in \AA{}. 
    (c) Side view of the Ge(111)-{\hskip 0pt}(3$\times$3)-Sn 
    reconstruction. Displacements relative to the room temperature phase 
    shown in (b) are given in \AA{}. 
    For clarity the displacements of symmetry equivalent atoms are only 
    shown once. 
    The atoms are marked with the same labels as in 
    Table~\ref{tab:abs_pos}. 
    }
  \label{fig:models}
\end{figure}

\begin{figure}[htb]
  \caption{(a) Comparison between the fractional order rods from both phases 
    The solid lines are calculated using the parameters 
    for the room temperature phase and 
    the data-points for this phase are marked by asterisks. 
    The corresponding curves for the low temperature phase are indicated by 
    dashed lines and squares. 
    Error bars have been omitted for clarity. 
    Close inspection reveals clear differences in intensity 
    for some rods (e.g. (2/3,8/3,$l$)), whereas other rods 
    are nearly identical for both phases (viz.\ (4/3,1/3,$l$)). 
    This proves that the 
    rods specific to the smaller room temperature unit-cell are very sensitive 
    to the structural changes upon cooling.
    (b) Fractional order rods unique to the (3$\times$3) surface 
    reconstruction.}
  \label{fig:rods}
\end{figure}
\vfill

\begin{table}[htb]
  \begin{center}
    \begin{tabular}[htb]{l|r|r|c}
      atom & pos.\ $\sqrt{3}$ &
      pos.\ (3$\times$3)  & dev.\ [\AA{}]\\\hline
      a  & (0.333,0.667, 0.563)& (0.333,0.667, 0.579) &            \\\hline
      b  &                     & (2.333,1.667, 0.569) &            \\\hline
      c  &                     & (1.333,2.667, 0.653) &            \\\hline
      1  & (0.024,0.048,-0.003)& (0.023,0.047, 0.020) &  0.17/0.17 \\\hline
      1' &                     & (2.026,1.052, 0.022) &       0.19 \\\hline
      1''&                     & (1.027,2.054, 0.056) &       0.26 \\\hline
      2  &(-0.667,0.667,-0.202)& (2.322,0.657,-0.171) &  0.16/0.26 \\\hline
      3  & (0.333,0.667,-0.364)& (0.333,0.667,-0.341) &  0.37/0.30 \\\hline
      3' &                     & (2.333,1.667,-0.364) &       0.37 \\\hline
      3''&                     & (1.333,2.667,-0.361) &       0.36 \\\hline
      4  &(-0.667,0.667,-0.971)& (2.334,0.663,-0.948) &  0.09/0.17 \\\hline
      5  & (0.333,0.667,-1.090)& (0.333,0.667,-1.069) &  0.29/0.23 \\\hline
      5' &                     & (2.333,1.667,-1.095) &       0.31 \\\hline
      5''&                     & (1.333,2.667,-1.084) &       0.28 \\\hline
      6  & (0.677,1.353,-1.257)& (0.681,1.362,-1.251) &  0.07/0.10 \\\hline
      6' &                     & (2.675,2.349,-1.240) &       0.06 \\\hline
      6''&                     & (1.674,0.349,-1.230) &       0.08 \\\hline
      7  & (0.671,1.343,-2.000)& (0.671,1.341,-1.995) &  0.03/0.03 \\\hline
      7' &                     & (2.673,1.327,-1.996) &       0.05 \\\hline
      7''&                     & (1.671,0.342,-1.993) &       0.04 \\\hline
      8  & (0.003,0.006,-2.253)& (0.006,0.012,-2.247) &  0.02/0.04 \\\hline
      8' &                     & (2.004,1.008,-2.245) &       0.03 \\\hline
      8''&                     & (1.000,0.000,-2.241) &       0.03 \\
     \end{tabular}
    \caption{The atomic positions in the room temperature 
      Ge(111)-{\hskip 0pt}($\sqrt{3}\times\!\sqrt{3}$)R30$^\circ$-Sn 
      and low temperature (3$\times$3) phases 
      and the deviations from ideal bulk-like positions in \AA{}. 
      The labels refer to Fig.~\ref{fig:models}. 
      For symmetry equivalent atoms only one 
      position is given. The estimated uncertainty 
      of the coordinates is about 0.02~\AA{}. 
      Isotropic atomic displacement parameters with amplitudes of 
      0.14~\AA{} (Sn), 0.12~\AA{} (four nearest-neighbor Ge atoms) and 
      0.09~\AA{} (Ge) were determined for the room temperature phase and 
      amplitudes of 
      0.41~\AA{} (Sn atom `c' in $z$-direction), 
      0.04~\AA{} (Sn `c' in-plane, `a' and `b'), 
      0.15~\AA{} (nearest-neighbor Ge atoms `{1''}' and `{3''}'), 
      0.08~\AA{} (remaining  Ge atoms in the layers 1-6) and 
      0.02~\AA{} (bulk Ge) for the low temperature phase.}
    \label{tab:abs_pos}
  \end{center}
\end{table}
\vfill



~\vspace{1cm}

\begin{center}
  \centerline{\epsfig{file=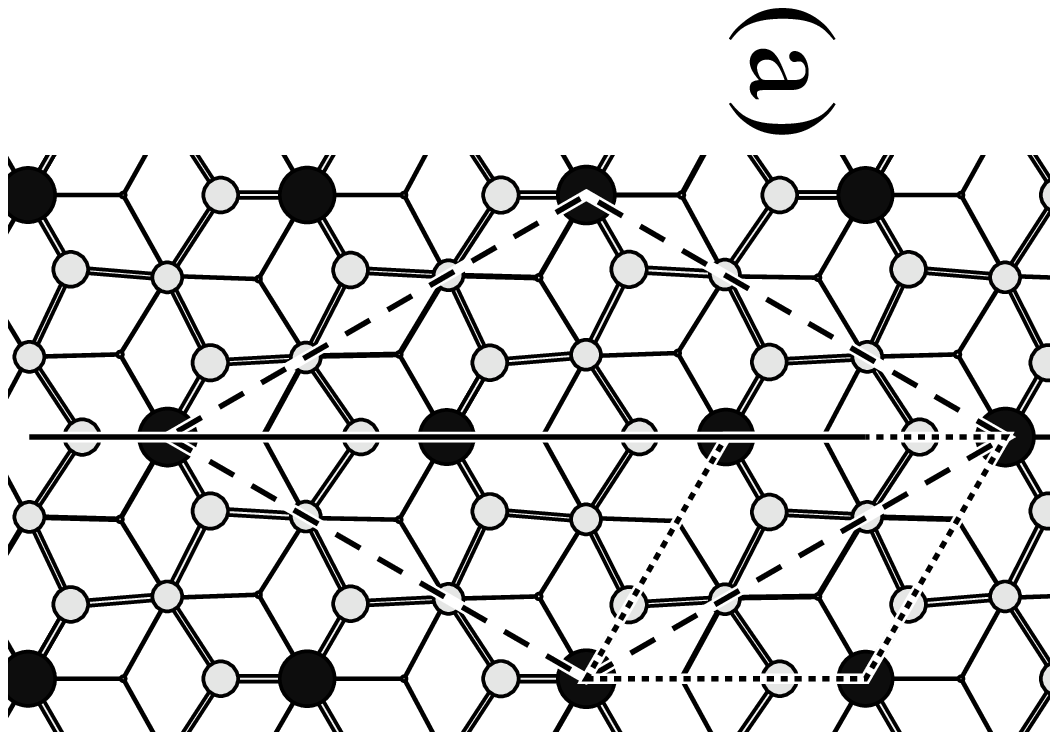,width=4.0cm,angle=90}
    \hspace{1mm}
    \epsfxsize=3.5cm
    \epsffile{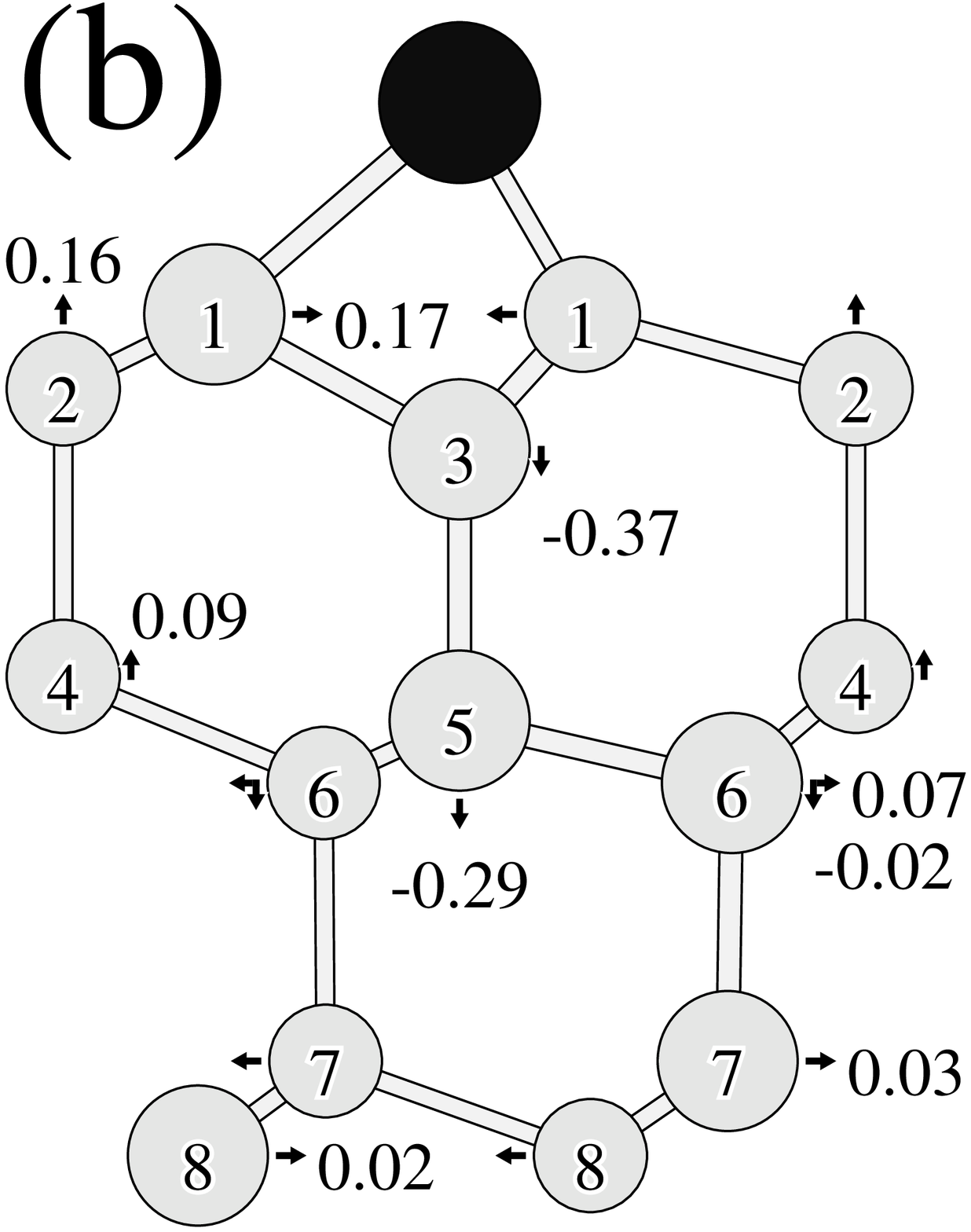}}

  \epsfxsize=8.5cm
  \epsffile{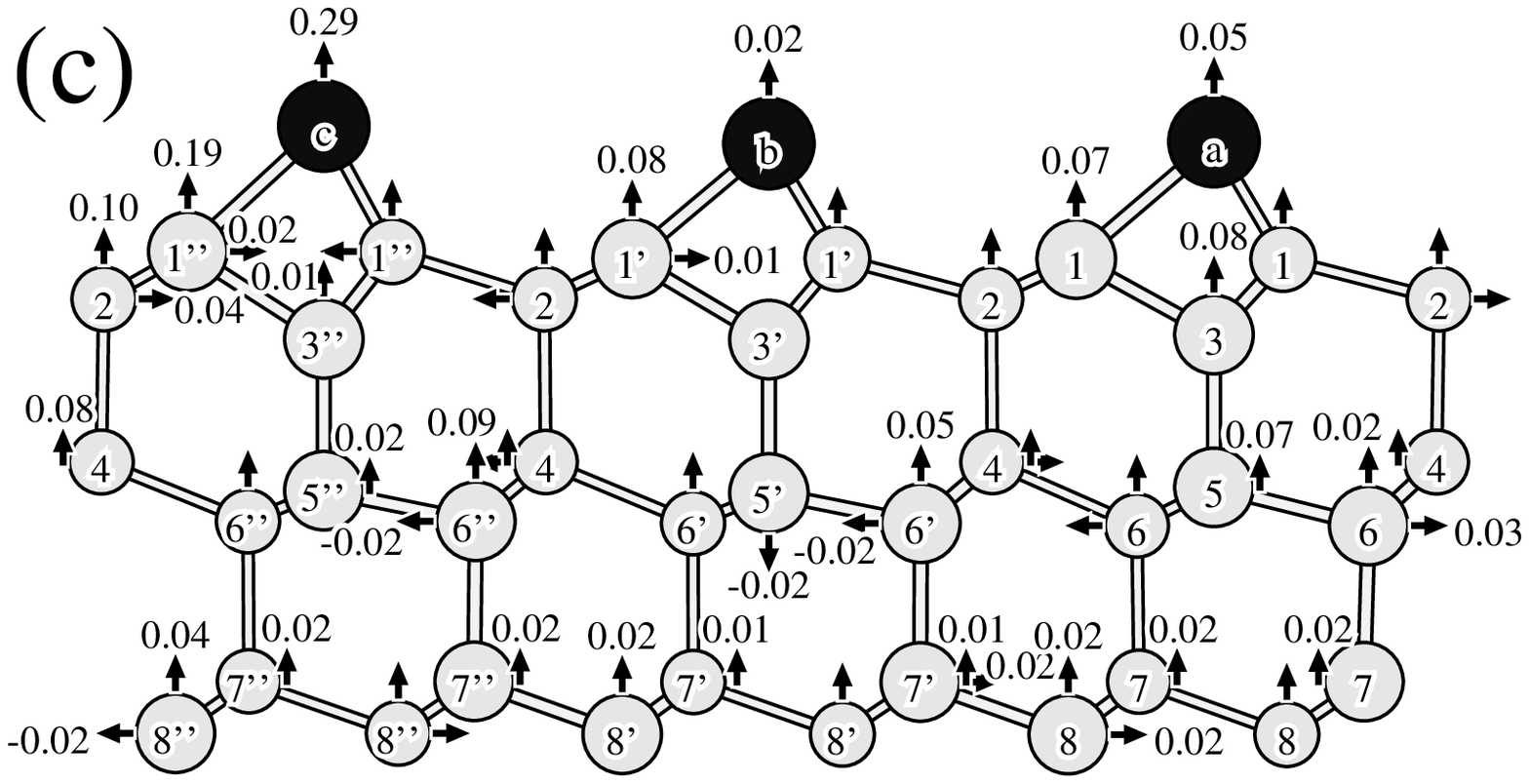}
  \vspace{1cm}

  {\Large Fig.~\ref{fig:models}}
\end{center}
\vfill


\begin{center}
\noindent {\Large (a)}
  \epsfxsize=6cm
  \epsffile{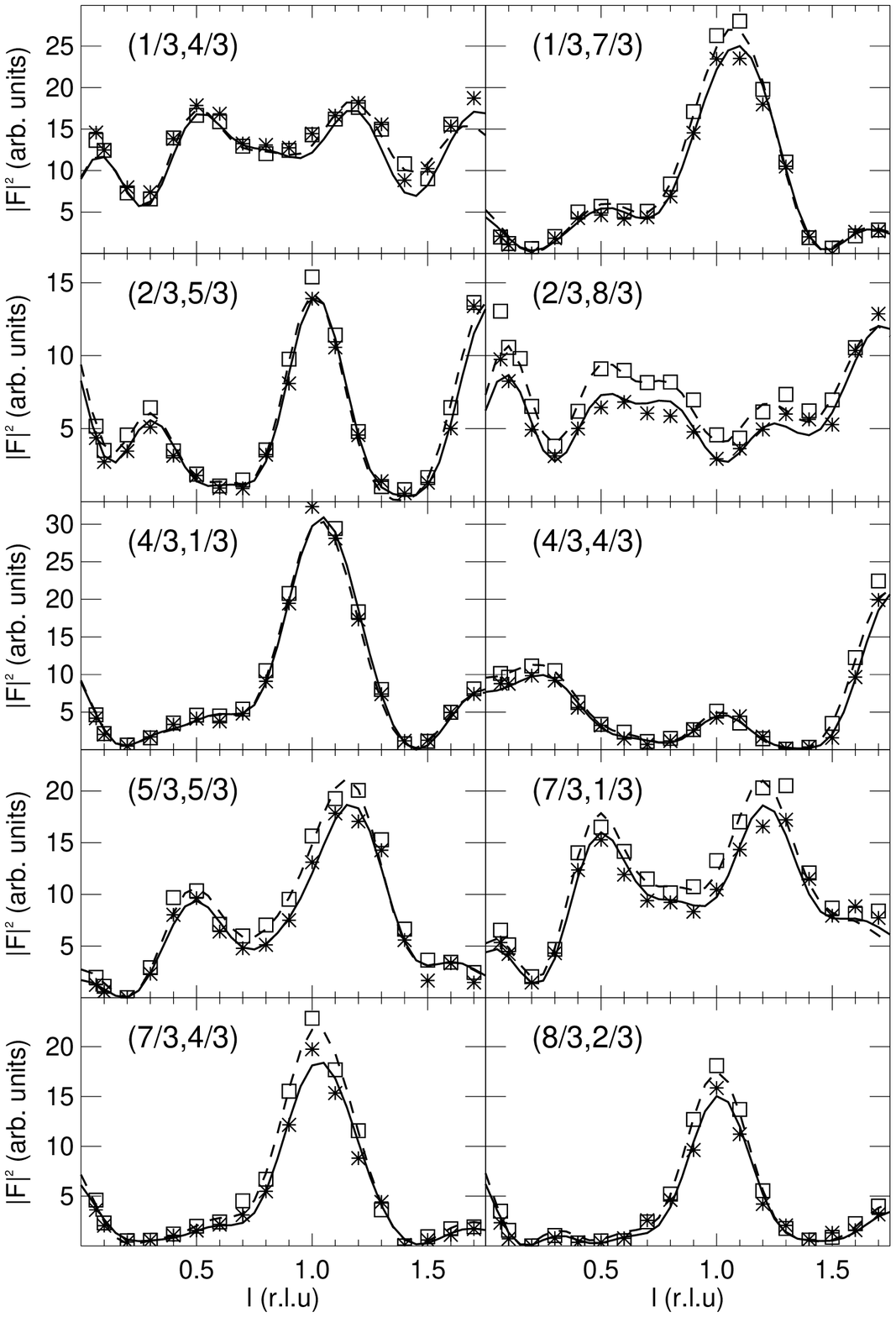}
\end{center}
\begin{center}
\noindent {\Large (b)}
  \epsfxsize=6cm
  \epsffile{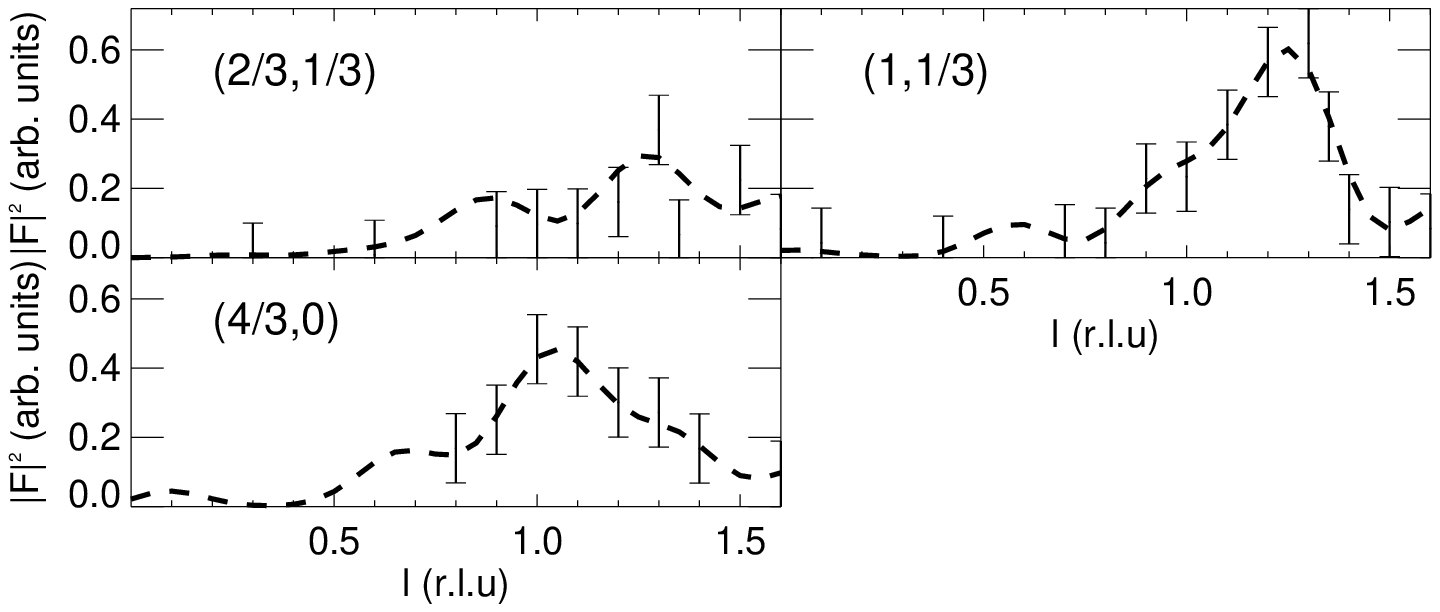}
  \vspace{1cm}

  {\Large Fig.~\ref{fig:rods}}
\end{center}

\end{multicols}
\end{document}